\documentclass[11pt]{article}
\usepackage{epsfig}
\newcommand{\be}{\begin{equation}}
\newcommand{\ee}{\end{equation}}
\newcommand{\ba}{\begin{eqnarray}}
\newcommand{\ea}{\end{eqnarray}}
\def\L{{\cal L}}

\newcommand{\vev}[1]{\left\langle #1 \right\rangle}

\newcommand{\psibar}{\overline{\psi}}
\newcommand{\dslash}{\hbox{$\partial$\kern-0.5em\raise0.3ex\hbox{/}}}
\def\slash#1{\hbox{$#1$\kern-0.5em\raise0.3ex\hbox{/}}}
\newcommand{\lr}[1]{\stackrel{\leftrightarrow}{\partial_{#1}}}
\begin{document}
\rightline{KOBE-TH-00-01}
\vspace{.5cm}
\begin{center}
{\LARGE QED out of matter}\\
\vspace{1cm} 
Hidenori SONODA\footnote{E-mail: sonoda@phys.sci.kobe-u.ac.jp}\\ 
\vspace{.2cm}
Physics Department, Kobe University, Kobe 657-8501, Japan\\
\vspace{.2cm} 
February 2000\\
\vspace{.2cm}
\end{center}
\vspace{.3cm}
\begin{abstract}
The Wilsonian renormalization group implies that an arbitrary four
dimensional field theory with an ultraviolet cutoff is equivalent to a
theory which is renormalizable by power counting at energy scales much
below the cutoff.  This applies to any theory including those with
non-renormalizable interactions as long as we fine-tune the mass
parameters.  We analyze two simple models with current-current
interactions but without elementary gauge fields from this viewpoint.
We show how to tune the parameters of the models so that they become
equivalent to QED at energies much below the cutoff.
\end{abstract}

\newpage
\section{Introduction}

There seem to be two kinds of field theories: one good and the other
bad.  Good theories are renormalizable theories, such as $\phi^4$
theory, QED, and QCD, which are well defined at all energy scales, and
for which everything can be calculated in terms of a finite number of
parameters.  Bad theories are non-renormalizable theories, such as the
four-Fermi weak theory and non-linear sigma model, which are well
defined only below a finite cutoff energy (or momentum), and for which
more and more parameters must be introduced as the order of
approximations increases.  Theories can be classified using the standard
power counting rule: if the lagrangian of a theory contains only fields
with dimension four or less, it is renormalizable, and otherwise it is
non-renormalizable.  This classification into renormalizable and
non-renormalizable theories is so simple and quite popular, but we know
it is wrong.

The physical meaning of renormalization and renormalizability cannot be
understood without the renormalization group (to be abbreviated as RG)
introduced by K.~G.~Wilson.\cite{WK} Using the renormalization group we
also distinguish two classes of theories, but this time one is the class
of $\phi^4$-like theories, and the other is that of QCD-like theories.

A $\phi^4$-like theory or a non-asymptotic free theory is characterized
by an IR fixed point of the renormalization group.  If the relevant (or
mass) parameters are taken to zero, the theory approaches the IR fixed
point along a RG trajectory.  With non-vanishing masses, the theory is
driven away from the fixed point.  The effects of irrelevant and
marginal parameters become smaller along a RG trajectory: the effect of
an irrelevant parameter, which is of order $1$ at the cutoff energy
scale $\Lambda$, is suppressed by a positive power of ${\mu \over
\Lambda}$ at an arbitrary low energy scale $\mu$, but the effect of a
marginal parameter of order $1$ at $\Lambda$ is only suppressed as ${1
\over \ln {\Lambda \over \mu}}$ at $\mu$.  We cannot take a true
continuum limit of this kind of theories; as we take $\Lambda \to
\infty$, the effects of marginal parameters vanish, and we are left with
a free massive theory.\footnote{We assume that the fixed point is a free
massless theory.}  This is called ``triviality'' in the literature.
(See Fig.~1.)

In contrast the other class of theories, QCD-like or asymptotic free
theories, admit a true continuum limit.  The fixed point of a QCD-like
theory is an UV fixed point.  We take the relevant parameters to zero as
we take the continuum limit.  The continuum limit obtained this way is
well-defined at all energy scales, and it is parameterized by mass and
relevant marginal parameters which drive the theory away from the fixed
point along the RG trajectories.  With a cutoff large but finite, the
theory differs from the continuum limit due to the effects of irrelevant
parameters, which are suppressed by positive powers of ${\mu \over
\Lambda}$.

\begin{center}
\epsfig{file=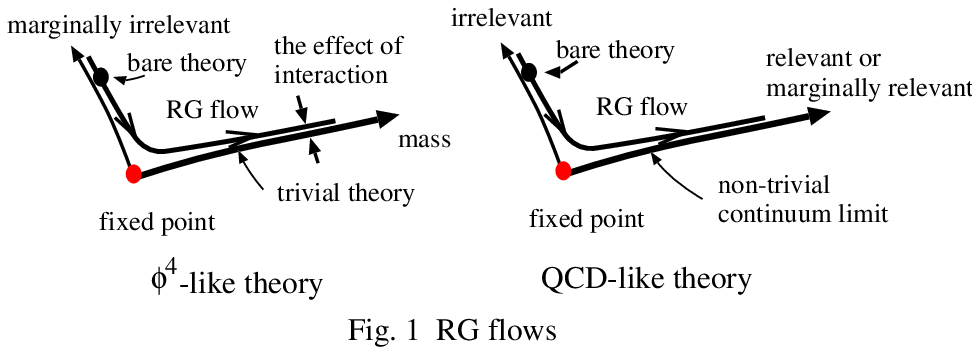, height=4cm}
\end{center}

Thus, any theory has good low energy behaviors as long as we fine-tune
the mass parameters.  Fine-tuning is necessary, since the natural mass
scale is $\Lambda$, and the mass parameter must be fine-tuned, typically
to the order of ${\mu^2 \over \Lambda^2}$, to attain a finite physical
mass much smaller than the cutoff.\footnote{In pure QCD we only need to
tune but not fine-tune the gauge coupling constant to zero, since it is
only marginally relevant.  With massive quarks, the quark masses must be
fine-tuned to zero to the order ${\mu \over \Lambda}$.}  A theory is
either QCD-like or $\phi^4$-like: if it is not QCD-like, it is
$\phi^4$-like, and with a large but finite cutoff $\Lambda$ and
fine-tuning of mass parameters the theory describes interactions of
order ${1 \over \ln {\Lambda \over \mu}}$ at a low energy scale $\mu$.

The above RG viewpoint was clearly stated in the first of
refs.~\cite{NJL} regarding the equivalence between the
non-renormalizable Nambu-Jona-Lasinio (NJL) model and the renormalizable
Yukawa theory.  Despite their difference in appearance, the two theories
are both $\phi^4$-like, and they describe the same physics if we ignore
the irrelevant differences suppressed by positive powers of ${\mu \over
\Lambda}$.  Similarly, the $O(N)$ linear- and non-linear sigma models
are both $\phi^4$-like, and they are equivalent up to differences
suppressed by positive powers of ${\mu \over \Lambda}$.\cite{ZJ} As
these two concrete examples show, the appearance of a theory at the
cutoff scale is misleading.  It is the fixed point of the RG which
dictates the low energy behaviors of a theory.

The purpose of the present paper is to extend the work of
refs.~\cite{NJL} and apply the above RG viewpoint to explore the
possibility of constructing gauge theories using apparently
non-renormalizable lagrangians without elementary gauge fields.  The
dynamical generation of gauge symmetries have already been discussed in
the literature.  Immediately after the work on the NJL model (the first
of refs.~\cite{NJL}), purely fermionic construction of QCD was attempted
in ref.~\cite{HH}.  Even earlier, in generalizing the idea of non-linear
realizations of symmetries, dynamical generation of gauge symmetries,
called hidden local symmetries, was shown to be possible.\cite{BKY} A
non-perturbative study of QCD constructed as an induced gauge theory has
been also given in ref.~\cite{KM}.

Our work differs from these earlier works in two aspects: first we will
study $\phi^4$-like gauge theories from the RG viewpoint given above.
We are not introducing new theories.  Rather we are showing that certain
non-renormalizable theories, often perceived as undesirable, are really
the same as perturbatively renormalizable gauge theories.  Second we
give a detailed analysis of the Ward identities. A renormalizable theory
with a vector field is not necessarily a gauge theory.  To be a gauge
theory, Ward identities must be satisfied.  We will show it possible to
tune marginally irrelevant parameters to satisfy the necessary Ward
identities.  In this paper we only consider two theories with abelian
gauge symmetries.

A comment is in order regarding the use of $1/N$ expansions in our work.
We emphasize that all our results are supported by the RG viewpoint
given above, and that they are valid for any $N$ starting from $1$.  We
use the $1/N$ expansions not to study the theories non-perturbatively.
We are only interested in perturbation theory with respect to small
coupling constants such as the fine structure constant and scalar
self-coupling.  For $\phi^4$-like theories, the so-called
non-perturbative effects are all cutoff dependent: their contributions
are suppressed by positive powers of ${\mu \over \Lambda}$ which we
ignore in our study of non-renormalizable theories.  Here we use the ${1
\over N}$ expansions to get a small coupling constant of order ${1 \over
\ln {\Lambda \over \mu}}$ from loop corrections.  Na\"{\i}ve
perturbative expansions in powers of a bare coupling does not work,
since the coupling is of order $1$.

This paper is organized as follows.  In sect.~2 we review the
equivalence of the $O(N)$ non-linear sigma model with the linear sigma
model at energies much below the cutoff.  This is to remind the reader
how misleading the appearance of a lagrangian can be and to emphasize
the usefulness of the $1/N$ expansions.  In sect.~3 we study a
non-renormalizable fermionic model with a current-current interaction
and show its equivalence to the standard (massive) QED.  This is
followed by the study of a little more non-trivial scalar model which
also has a current-current interaction in sect.~4.  The model is shown
to be equivalent to the (massive) scalar QED.  Finally, the paper is
concluded in sect.~5.  For the convenience of the reader, we summarize
the Ward identities of QED and scalar QED in appendix A.  A table of
integrals with a momentum cutoff is given in appendix B.

Throughout the entire paper we will work in the four dimensional
euclidean space.  Our convention is that the weight of a euclidean
functional integral is given by $\exp [ - S] = \exp [ - \int d^4 x~\L]$
where $S$ is a euclidean action, and $\L$ is a euclidean action
density.\footnote{We call it an action density rather than a lagrangian
density to avoid potential confusion about signs.}

\section{Review of the $O(N)$ non-linear sigma model}

We begin with a brief review of the equivalence between the $O(N)$
linear and non-linear sigma models.\cite{ZJ} We wish to explain
quantitatively how the apparently non-renormalizable non-linear sigma
model can be physically equivalent to the perturbatively renormalizable
linear sigma model.  We first summarize the relevant results on the
linear sigma model.\cite{linear} The model is defined for $N$ real
scalar fields $\phi^I$ ($I=1,...,N$) by the following action density
\be
\L = {1 \over 2} \partial_\mu \phi^I \partial_\mu \phi^I +
{m^2 \over 2} \phi^I \phi^I + {\lambda \over 8 N} \left( \phi^I \phi^I
\right)^2 
\ee
with a momentum cutoff $\Lambda_{\rm L}$.  To leading order in ${1 \over
N}$, the renormalized squared mass $m_r^2$ and the renormalized
self-coupling $\lambda_r$ are obtained as\footnote{The terms of order
${m_r^2 \over \Lambda_{\rm L}^2}$ and less are ignored.}
\ba
m^2 &=& m_r^2 + {\lambda \over 2 (4\pi)^2} \left( - \Lambda_{\rm L}^2 +
m_r^2 \ln {\Lambda_{\rm L}^2 \over m_r^2}\right)\\ {1 \over \lambda_r}
&=& {1 \over \lambda} + {1 \over 2(4\pi)^2} \ln {\Lambda_{\rm L}^2 \over
\mu^2} \equiv {1 \over 2(4\pi)^2} \ln {\Lambda_0^2 \over
\mu^2} \label{landau}
\ea 
where $\mu$ is an arbitrary renormalization scale.  The correlation
functions of $\phi^I$ are made UV finite in terms of $m_r^2$ and
$\lambda_r$.  This is the standard renormalization.  In
Eq.~(\ref{landau}) we have introduced the Landau scale $\Lambda_0$ at
which the bare coupling $\lambda$ diverges for a given $\lambda_r$.  The
Landau scale gives the largest energy scale beyond which the theory is
not defined.  The linear sigma model is thus characterized uniquely by
two parameters: $m_r^2$ and $\Lambda_0$.

Let us now take a look at the non-linear sigma model.  It is defined by
the action density
\be
\L = N \left[ {v^2 \over 2} \partial_\mu \Phi^I \partial_\mu \Phi^I + {a
\over 8} \left( \partial_\mu \Phi^I \partial_\mu \Phi^I \right)^2 \right]
\label{NL}
\ee
with a momentum cutoff $\Lambda_{\rm NL}$.  Due to the non-linear
constraint
\be
\Phi^I \Phi^I = 1,
\ee
the theory is not renormalizable by the usual power counting.  

Some comments on the four-derivative term in Eq.~(\ref{NL}) are in
order.  At first sight it seems totally irrelevant; if we expand the
action density na\"{\i}vely in terms of the unconstrained fields $\pi^i
= \sqrt{N} v \Phi^i$ ($i=1,...,N-1$), the four-derivative term gives
rise to an interaction term suppressed by ${1 \over v^2}$ which is of
order ${1 \over \Lambda_{\rm NL}^2}$.  The four-derivative term is
indeed irrelevant but not for that reason; it can give marginal
contributions, i.e., contributions not suppressed by inverse powers of
the momentum cutoff, at low energies.  Here, the role of $a$ is merely
to rescale the momentum cutoff, and therefore it is physically
irrelevant.

It is the existence of a critical value $v_c^2$ which is the key to the
renormalizability of the non-linear model and its equivalence to the
linear model.  At the critical point $v^2 = v_c^2$ the theory becomes a
theory of $N$ free massless scalars.  For $v^2 < v_c^2$ the larger
fluctuations of the fields are encouraged, and the $O(N)$ symmetry is
fully restored.  For $v^2 > v_c^2$, however, the fluctuations are
discouraged, and the $O(N)$ is spontaneously broken to $O(N-1)$.  By
fine-tuning the parameter $v^2$ near $v_c^2$, the non-linear sigma model
gives the same physics as the linear sigma model; all the differences
are suppressed by positive powers of ${\mu^2 \over \Lambda_{\rm NL}^2}$
where $\mu$ is an arbitrary but finite renormalization scale.

The large $N$ calculations for the non-linear model is well
known.\cite{ZJ} For simplicity, we restrict ourselves to the symmetric
phase.  To leading order in ${1 \over N}$, the critical value $v_c^2$ is
given by
\be
{v_c^2 \over z} \equiv {\Lambda_{\rm NL}^2 \over (4 \pi)^2},
\ee
where
\be
z \equiv 1 - (4 \pi)^2 {a \over 4}.
\ee
We take $a < {4 \over (4\pi)^2}$ so that $z > 0$.

By straightforward calculations, we can verify that the non-linear model
is equivalent to the linear sigma model with renormalized squared mass
$m_r^2$ and self-coupling $\lambda_r$, if we choose $v^2$ by
\be
{v^2 - v_c^2 \over z} = - {m_r^2 \over (4 \pi)^2} \ln {\Lambda_{\rm
NL}^2 \over m_r^2} < 0
\ee
and choose the cutoff $\Lambda_{\rm NL}$ by
\be
\ln {\Lambda_{\rm NL}^2 \over \Lambda_0^2} = {2(4\pi)^2} {a
\over 4}
\ee
where $\Lambda_0$ is related to $\lambda_r$ by Eq.~(\ref{landau}).  We
see that the constant $a$ merely changes the ratio of $\Lambda_{\rm NL}$
to $\Lambda_0$ by a finite amount.

A similar analysis can be given to the Nambu-Jona-Lasinio
model.\cite{NJL} The Nambu-Jona-Lasinio model with a non-renormalizable
Fermi interaction is equivalent to a perturbatively renormalizable
Yukawa theory if we ignore contributions suppressed by negative powers
of the momentum cutoff.

\section{QED with electrons}

In this and next sections we analyze models with non-renormalizable
current-current interactions.  We first consider a purely fermionic
theory defined by the following action density\footnote{Our convention
for euclidean fermionic fields might differ somewhat from the standard
convention.  Replace $\psibar$ by $i \psibar$ to get a more familiar
kinetic term $\psibar (\dslash - M)\psi$.  The hermitian gamma matrices
$\gamma_\mu$ satisfy the Clifford algebra $\{\gamma_\mu, \gamma_\nu \} =
2 \delta_{\mu\nu}$ as usual.}
\be
\L = \psibar^I \left( {1 \over i} \dslash + i M \right)\psi^I -
 {1 \over 2 N v^2} J_\mu J_\mu, \label{lagrangian}
\ee
where $I$ runs from $1$ to $N$, and the current $J_\mu$ is defined by
\be
J_\mu \equiv \psibar^I \gamma_\mu \psi^I.\label{current}
\ee
To define a theory we introduce a momentum cutoff $\Lambda$.  It is
essential to use a momentum cutoff as opposed to the dimensional
regularization.  We are interested in the dependence of the theory on
the UV cutoff, and the dimensional regularization is not suitable for
this purpose, since it automatically gives the limit of an infinite
cutoff.

It turns out that the theory defined by the action density
(\ref{lagrangian}) and the current (\ref{current}) is missing one
marginal parameter.\footnote{Similarly, in ref.~\cite{NJL} a dimension
eight field was introduced to the action density to account for a
missing marginal parameter in the na\"{\i}ve NJL model.}  Instead of the
current given by Eq.~(\ref{current}), we consider a more general
\be
J_\mu \equiv \psibar^I \gamma_\mu \psi^I + {h \over \Lambda^2} ~\psibar^I
\overleftarrow{\dslash} \gamma_\mu \dslash \psi^I
\label{general_current}
\ee
The momentum cutoff does not respect gauge invariance, and we need to
adjust the coefficient $h$ for the Ward identities.\footnote{See
Appendix A for a summary of the Ward identities for QED.}

It is important to observe that the action density $\L$ is invariant
under the charge conjugation ${\cal C}$ defined by
\be
\psi \to C \psibar^T, \quad
\psibar \to - \psi^T C^{-1}
\ee
where the four-by-four matrix $C$ satisfies
\be
C^{-1} \gamma_\mu C = - \gamma_\mu^T.
\ee
The current $J_\mu$ is odd under ${\cal C}$.

It is not necessary but helps our calculations to introduce a vector
auxiliary field $A_\mu$.  We then rewrite the action density as
\ba
\L &=& \psibar^I \left( {1 \over i} \dslash + i M \right)\psi^I - {1
\over 2 N v^2} J_\mu J_\mu + {1 \over 2} \left(v A_\mu + {1 \over
\sqrt{N} v} J_\mu \right)^2 \nonumber \\ &=& \psibar^I \left( {1 \over
i} \dslash + i M \right) \psi^I + {1 \over 2} v^2 A_\mu^2\nonumber \\
&&\qquad + {1 \over \sqrt{N}} A_\mu \left( \psibar^I \gamma_\mu \psi^I +
{h \over \Lambda^2}~ \psibar^I \overleftarrow{\dslash} \gamma_\mu \dslash
\psi^I \right) \label{Lfermi}
\ea

What do we expect at energy scales much below the cutoff $\Lambda$?  We
ignore the effects suppressed by the inverse powers of the cutoff
$\Lambda$ but keep those only suppressed by negative powers of the
logarithm of $\Lambda$.  If we fine-tune the mass parameter $v$ so that
the mass scale of the theory remains UV finite, we expect that the above
theory becomes equivalent to a theory which is renormalizable by power
counting.  The smallest renormalizable theory with fermions $\psi^I$ and
a real vector field $A_\mu$ is given by the following action density:
\be
\L_{ren} = {1 \over 4} F_{\mu\nu}^2 + {1 \over 2 \xi_0} (\partial_\mu
A_\mu)^2 + {m_0^2 \over 2} A_\mu^2 + {\lambda_0 \over 8N} (A_\mu^2)^2
+ \psibar^I \left( {1 \over i} \dslash + {e_0 \over \sqrt{N}} \slash{A} +
i M \right) \psi^I \label{Lren}
\ee
where we have imposed the $\cal C$ invariance.  At tree level the Ward
identity demands that the parameter $\lambda_0$ vanish.  But
renormalizability alone allows an arbitrary $\lambda_0$.

To leading order in ${1 \over N}$ the fermion-photon vertex receives no
radiative correction, and we only need to calculate the one-loop
contributions to the two- and four-point proper vertices of the photon
field $A_\mu$.  The three-point vertex vanishes due to the $\cal C$
invariance.  If the theories defined by (\ref{Lfermi}) and (\ref{Lren})
have the same two- and four-point vertices, the two theories are
equivalent, since any higher point functions can be constructed out of
the fermion-photon vertex and the photon two- and four-point vertices
independently of the cutoff $\Lambda$.

The one-loop calculations with a momentum cutoff $\Lambda$ are
straightforward, and we only write the results here.  The inverse
propagator is calculated from the top one-loop diagram in Fig.~2 as
follows:
\ba
&&\Pi_{\alpha\beta} (k^2) = {1 \over e^2} \Bigg[ m_\gamma^2
\delta_{\alpha\beta} + {1 \over \xi} k_\alpha k_\beta \\ && + (k^2
\delta_{\alpha\beta} - k_\alpha k_\beta) \left( 1 - 8 {e^2 \over (4\pi)^2}
\int_0^1 dx~x(1-x) \ln {M^2 + x(1-x) k^2 \over \mu^2}
\right)\Bigg]\nonumber
\ea
where we define
\ba
{(4 \pi)^2 \over e^2} &\equiv& {4 \over 3} \ln {\Lambda^2 \over \mu^2} - 1 
+ {4 \over 3} h - {1 \over 9} h^2 \\
{(4 \pi)^2 \over e^2} m_\gamma^2 &\equiv& (4 \pi)^2 v^2 +
\left(-2+4 h - {2 \over 3} h^2\right) \Lambda^2 + (2-12h) M^2 \label{mass}\\
{(4 \pi)^2 \over e^2} {1 \over \xi} &\equiv& {1 \over 3} (1 + 4 h - h^2)
\ea
Eq.~(\ref{mass}) implies that we must fine-tune the squared mass
parameter $v^2$ so that the squared mass $m_\gamma^2$ of the photon is
finite and positive.  This is fine-tuning as opposed to tuning, since
${v^2 \over \Lambda^2}$ must be tuned to an order $1$ quantity to the
accuracy of ${M^2 \over \Lambda^2}$.  With this fine-tuning, the above
photon two-point function is identical to the one in the massive QED
with the running gauge coupling constant $e$, photon mass $m_\gamma$,
and gauge fixing parameter $\xi$.  The mass parameter $\mu$ is an
arbitrary renormalization scale.
\begin{center}
\epsfig{file=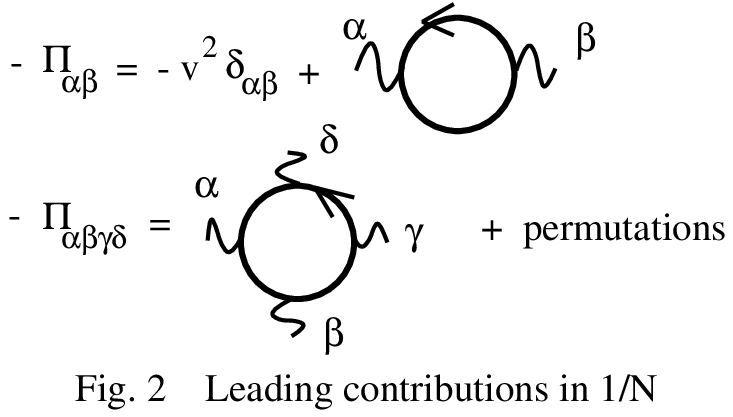, height=3.5cm}
\end{center}

The four-point function at zero external momenta is obtained from the
bottom one-loop graphs in Fig.~2, and it depends on $h$:
\be
\Pi_{\alpha\beta\gamma\delta} = {1 \over N} {1 \over (4 \pi)^2} \left(
\delta_{\alpha\beta} \delta_{\gamma\delta} + \delta_{\alpha\gamma}
\delta_{\beta\delta} + \delta_{\alpha\delta} \delta_{\beta\gamma}
\right) \left( {4 \over 3} - 16 h + 24 h^2 - {16 \over 3} h^3 \right).
\ee
For the theory to be the massive QED, the Ward identity for the
four-point function must be satisfied.  Hence, we must choose the
parameter $h$ such that the above four-point function vanishes:
\be
{4 \over 3} - 16 h + 24 h^2 - {16 \over 3} h^3 = 0.
\ee
This equation has three real roots.  We can choose, for example, $h
\simeq 0.097$.  We note that the choice of $h$ is a tuning, but not a
fine-tuning; it must be tuned relative to order $1$, but not ${\mu^2
\over \Lambda^2}$.

The result corresponding to $m_\gamma = 0$ can be obtained from the
induced QED which is defined by the action density (\ref{Lfermi}) with a
specific choice $v=h=0$.\footnote{In addition we must use a cutoff which
respects the gauge invariance.  Otherwise the electron loops will not be
gauge invariant.} We have chosen to study the more general
(\ref{Lfermi}) because our interest is to verify the equivalence of the
non-renormalizable theory to the renormalizable theory which is defined
by Eq.~(\ref{Lren}) with arbitrary parameters $m_0^2, \lambda_0$.

We have chosen the interaction to have the current-current form, but it
was not necessary.  Instead of modifying the current by a dimension $5$
term proportional to $h$, we could have introduced a counterterm
\be
{\rm const} \times \left({1 \over N \Lambda^4} \left(\psibar^I
\gamma_\mu \psi^I\right)^2 \right)^2
\ee
in the action density.  This is how we introduce missing marginal
parameters for scalar QED in the next section.

Before closing this section, we make a remark on the sign of the
interaction term in the action density (\ref{lagrangian}).  Since the
current $J_\mu$ defined by Eq.~(\ref{general_current}) is a real field,
the current-current interaction term in (\ref{lagrangian}) is negative
definite.  We believe it does not invalidate the theory.  Our reasoning
goes as follows.  First we note that the theory defined by
(\ref{lagrangian}) is equivalent to the massive QED, which is a stable
theory, modulo irrelevant differences of order ${\mu^2 \over
\Lambda^2}$.  Therefore, if the theory is unstable, the effects of the
instability must be suppressed by positive powers of ${\mu^2 \over
\Lambda^2}$.  This suggests that a potential instability can only arise
from the large fluctuations of the fields, for example $A_\mu$ of order
$\Lambda$.  If this is the case, stability will be assured by redefining
the theory by the action density (\ref{Lfermi}) where the auxiliary
field $A_\mu$ is restricted within a finite range $|A_\mu| < \Lambda$.
The effects of this modification are suppressed by positive powers of
${\mu^2 \over \Lambda^2}$.

\section{Scalar QED}

The fermionic theory may be a little too simple.  It resembles the
induced QED too much.  If we had used a regularization which allows
shifts of momentum such as the dimensional regularization, the vacuum
polarization would have come out transverse, and the photon four-point
function would have vanished at zero external momenta.  The Ward
identities are then satisfied automatically.  Let us introduce a more
non-trivial example of a purely bosonic theory in this section.

The theory is defined by the following action density:
\ba
\L &=& \partial_\mu \phi^{I*} \partial_\mu \phi^I + m^2 \phi^{I*} \phi^I
+ {\lambda \over 4N} \left(\phi^{I*} \phi^I \right)^2 \nonumber \\
&& - {v^2 \over 2}
\left( {- i \over \sqrt{N} v^2} \phi^{I*}\lr{\mu} \phi^I\right)^2
+ \Delta \L,\label{Lscalar}
\ea
where the asterisk $*$ denotes complex conjugation, and
the counterterm $\Delta \L$ is defined by
\ba
\Delta \L &\equiv& {a\over N} (\phi^{I*} \phi^I) {1 \over 2} \left( {-i
\over \sqrt{N} v^2} \phi^{I*} \lr{\mu} \phi^I \right)^2 \nonumber\\
&& \quad + {b\over N (4
\pi)^2} {1 \over 8} \left\{\left( {-i \over \sqrt{N} v^2} \phi^{I*}
\lr{\mu} \phi^I \right)^2 \right\}^2.
\ea
We have introduced enough number of parameters so that the theory is
equivalent to the theory defined by the following renormalizable
action density:
\ba
\L_{ren} &=& \partial_\mu \phi^{I*} \partial_\mu \phi^I + m_0^2
\phi^{I*} \phi^I + {\lambda_0 \over 4N} (\phi^{I*} \phi^I)^2 \nonumber\\
&& + {1 \over 4} F_{\mu\nu}^2 + {1 \over 2 \xi_0} (\partial_\mu A_\mu)^2
+ {m_{\gamma,0}^2 \over 2} A_\mu^2 \nonumber \\ && + {e_0 \over
\sqrt{N}} A_\mu i \phi^{I*} \lr{\mu} \phi^I + {\gamma \over 2N}
\phi^{I*} \phi^I A_\mu^2 + {\delta \over 8 N} {1 \over (4\pi)^2}
(A_\mu^2)^2
\ea
Just as the Ward identities can determine the constants $\gamma$ and
$\delta$ uniquely, we will be able to fix the coefficients $a, b$ of the
counterterms by imposing the Ward identities.  We should note that
unlike the fermionic theory discussed in the previous section, the
interaction of this theory is not solely current-current type due to the 
counterterms.

To facilitate the ${1 \over N}$ expansions, we introduce scalar and
vector auxiliary fields $\alpha, A_\mu$ and rewrite the action density
as follows:
\ba
\L &=& \partial_\mu \phi^{I*} \partial_\mu \phi^I + m^2 \phi^{I*} \phi^I
+ {\lambda \over 4N} \left(\phi^{I*} \phi^I \right)^2 - {v^2 \over 2}
\left( {- i \over \sqrt{N} v^2} \phi^{I*}\lr{\mu}
\phi^I\right)^2\nonumber\\ && + {1 \over 2} \left( \alpha + i
\sqrt{\lambda \over N} \phi^{I*} \phi^I\right)^2 + {1 \over 2} \left( v
A_\mu + {i \over \sqrt{N} v} \phi^{I*} \lr{\mu} \phi^I \right)^2 +
\Delta \L\nonumber\\ &=& \partial_\mu \phi^{I*} \partial_\mu \phi^I + {1
\over 2} \alpha^2 + i \sqrt{\lambda \over N} ~\alpha~ \phi^{I*} \phi^I
\nonumber\\ && \quad + {v^2 \over 2} A_\mu^2 + {1 \over \sqrt{N}} A_\mu
i \phi^{I*} \lr{\mu} \phi^I + \Delta \L.
\ea
Unlike the fermionic theory of the previous section, this theory
does not reduce to an induced QED for $v=0$.

To leading order in ${1 \over N}$, we must renormalize the scalar mass
and self-coupling as
\ba
\Delta m^2 &\equiv& m_r^2 - m^2 = {\lambda \over (4\pi)^2} \left
( \Lambda^2 - m_r^2 \ln {\Lambda^2 \over m_r^2} \right)\\
{(4 \pi)^2 \over \lambda_r} &=& {(4 \pi)^2 \over \lambda} 
+ \ln {\Lambda^2 \over \mu^2} - 1
\ea
and shift the auxiliary field $\alpha$ by
\be
\alpha = - i \sqrt{N \over \lambda}~ \Delta m^2 + \sqrt{\lambda_r \over
\lambda}~ \delta \alpha,
\ee
where the shifted field $\delta \alpha$ has a vanishing expectation
value $\vev{\delta \alpha} = 0$.  To leading order in $1 \over N$, the
full propagator of the fluctuation $\delta \alpha$ is given by
\be
\vev{\widetilde{\delta \alpha} (k) \delta \alpha} = 1\Bigg/\left( 1 +
{\lambda_r \over (4 \pi)^2} \int_0^1 dx ~\ln {\mu^2 \over m_r^2 + x(1-x)
k^2} \right).
\ee

Before calculating the vertex functions involving the vector field
$A_\mu$, we note the implication of the equation of motion for $A_\mu$.
The action density is quadratic with respect to $A_\mu$, and the
equation of motion gives
\be
A_\mu = B_\mu \equiv {- i \over \sqrt{N} v^2} \phi^{I*} \lr{\mu} \phi^I
\label{EOM}
\ee
This implies that the composite field $B_\mu$ is an interpolating field
of the photon.  The calculation of the proper vertex of $A_\mu$ and
$B_\nu$ indeed gives
\be
\vev{\widetilde{A_\mu}(k) B_\nu} =
\delta_{\mu\nu} {1 \over v^2} \int_p {p^2 \over (p^2 + m_r^2)^2} \simeq
\delta_{\mu\nu} {1 \over v^2} {\Lambda^2 \over (4 \pi)^2}
\ee
where we have ignored the terms of order ${m_r^2 \over \Lambda^2}$.
(Fig.~3) This is $\delta_{\mu\nu}$ if we choose
\be
v^2 \simeq {\Lambda^2 \over (4 \pi)^2}. \label{vlambda}
\ee
We will see that Eq.~(\ref{vlambda}) is required by the fine-tuning of
the mass parameter.  Hence, the equation of motion (\ref{EOM}) implies
that the counterterm in the action density is equivalent to
\be
\Delta \L = {a \over 2 N} \left( \phi^{I*} \phi^I \right) A_\mu^2 + {b
\over 8 N} (A_\mu^2)^2
\ee
to leading order in ${1 \over N}$.  Therefore, $\Delta \L$ gives rise to
the vertices in Fig.~4.
\begin{center}
\epsfig{file=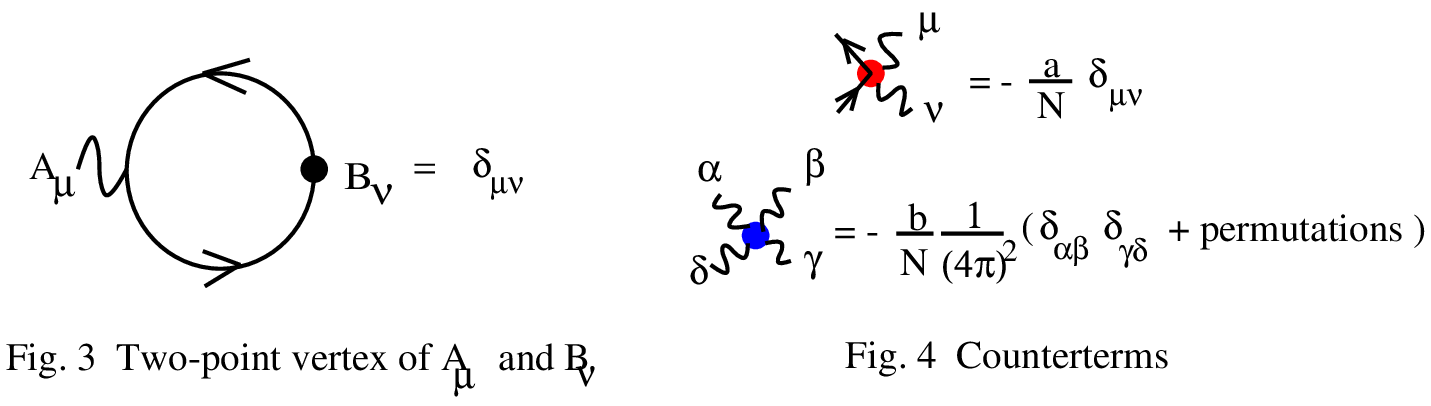, height=3cm}
\end{center}

Let us proceed with the two-point vertex of the photon.  To leading
order in ${1 \over N}$, we obtain
\ba
&&\Pi_{\alpha\beta} = {1 \over e^2} \Bigg[ m_\gamma^2
\delta_{\alpha\beta} + {1 \over \xi} k_\alpha k_\beta\\ && +
(k^2 \delta_{\alpha\beta} - k_\alpha k_\beta) \left( 1 - {e^2 \over (4
\pi)^2} \int_0^1 dx (1-2x)^2 \ln {m_r^2 + x(1-x) k^2 \over \mu^2}
\right) \Bigg],\nonumber
\ea
where the renormalized parameters are defined by
\ba
{(4\pi)^2 \over e^2} &\equiv& {1 \over 3} \ln {\Lambda^2 \over \mu^2} -
{1 \over 2} \\ m_\gamma^2 &\equiv& e^2 \left[ v^2 - {1 \over (4 \pi)^2}
\left( \Lambda^2 + m_r^2 - 2 m_r^2 \ln {\Lambda^2 \over m_r^2} \right)
\right] \label{photon_mass}\\ {1 \over \xi} &\equiv& - {e^2 \over (4\pi)^2} {1
\over 6}.
\ea
Here, $\mu$ is an arbitrary renormalization scale as usual.
Eq.~(\ref{photon_mass}) implies that we need to fine-tune the squared
mass parameter $v^2$ such that $0 < m_\gamma^2 \ll \Lambda^2$.

Now that we have identified the photon mass $m_\gamma$ and the gauge
coupling $e$, we wish to proceed with verifying the equivalence of our
theory with the scalar QED.  This will be done by checking three Ward
identities.  (See Appendix A for a summary of the Ward identities.)  The
first is the Ward identity for the three-point vertex of the scalar and
photon.  To leading order in ${1 \over N}$, the vertex receives no
radiative correction (Fig.~5), and the Ward identity is automatically
satisfied.
\begin{center}
\epsfig{file=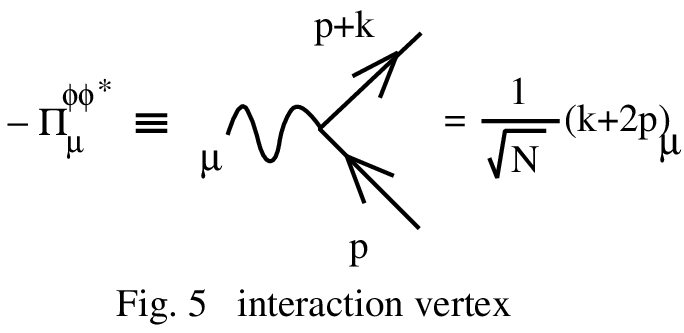, height=3cm}
\end{center}

Next we examine the scalar-scalar-photon-photon vertex.  To leading
order in ${1 \over N}$, it is given by the four diagrams in Fig.~6,
where we denote the propagator of $\delta \alpha$ by a broken line.
\begin{center}
\epsfig{file=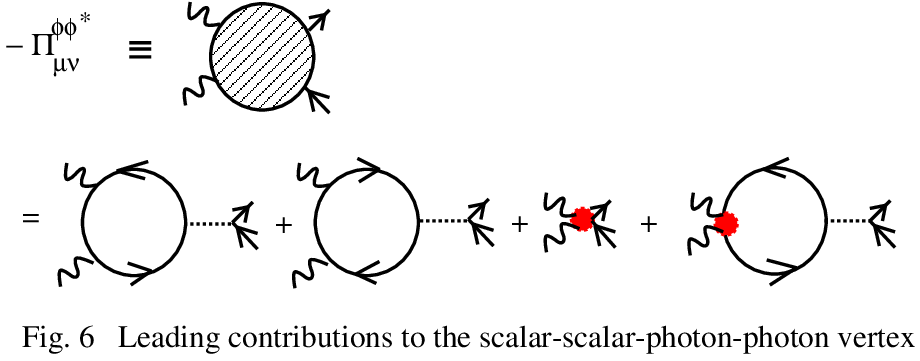, height=4cm}
\end{center}
The third and fourth terms involve the counterterm proportional to $a$.
The Ward identity demands that at zero external momenta this be given by
\be
- \Pi_{\mu\nu}^{\phi\phi^*} \Big|_{\rm zero~momenta} = - {2 \over N}
~\delta_{\mu\nu}.\label{AAss}
\ee
To leading order in ${1 \over N}$ we obtain
\be
- \Pi_{\mu\nu}^{\phi\phi^*} \Big|_{\rm zero~momenta} = - {2 \over
N}~\delta_{\mu\nu} {\ln {\Lambda^2 \over m_r^2} - 1 + {a \over 2} {(4
\pi)^2 \over \lambda} - {1 \over 2} \over
\ln {\Lambda^2 \over m_r^2} - 1 + {(4 \pi)^2 \over \lambda} }.
\ee
Therefore, we must choose $a$ as
\be
a = {\lambda \over (4 \pi)^2} + 2.\label{a}
\ee
This is an ordinary tuning, since $\lambda$ is a quantity of order $1$.

Finally we examine the four-photon vertex.  Many graphs contribute to
leading order in ${1 \over N}$ as in Fig.~7.
\begin{center}
\epsfig{file=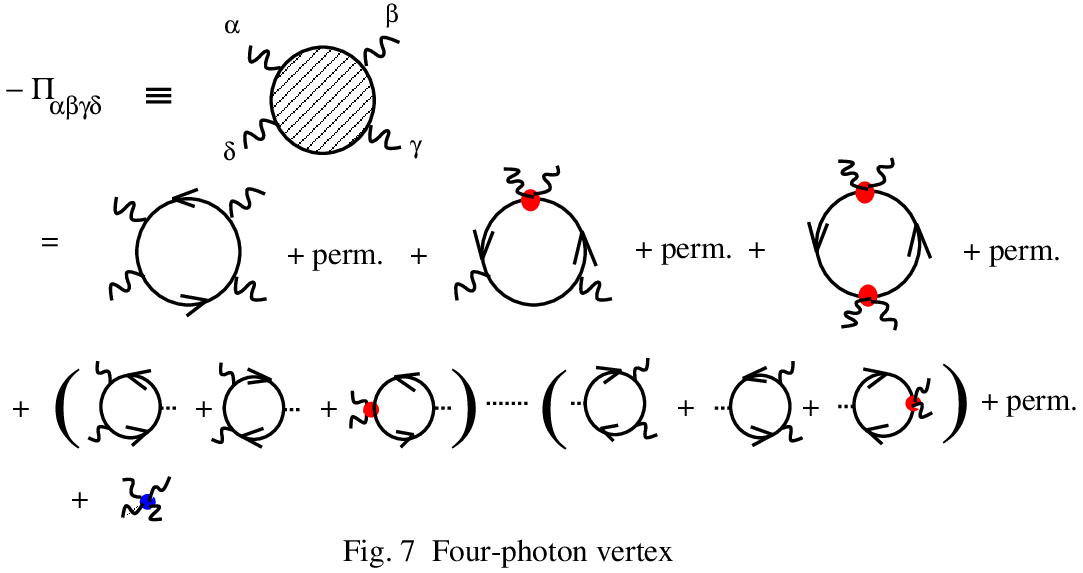, height=6cm}
\end{center}
The last term in Fig.~7 is the counterterm proportional to $b$.  The
Ward identity demands that the four-photon vertex vanish for zero
external momenta.  The calculation is straightforward but somewhat
involved.  We can use Eq.~(\ref{AAss}) to simplify the calculation.
The final result is given by
\be
- \Pi_{\alpha\beta\gamma\delta}\Big|_{\rm zero~momenta} = {1 \over N} {1
\over (4 \pi)^2} \left( \delta_{\alpha\beta} \delta_{\gamma\delta} +
\delta_{\alpha\gamma} \delta_{\beta\delta} + \delta_{\alpha\delta}
\delta_{\beta\gamma} \right) \left( - {4 \over 3} + a - b \right).
\ee
Therefore, the Ward identity demands
\be
b = {\lambda \over (4\pi)^2} + {2 \over 3}\label{b}
\ee
where we have used Eq.~(\ref{a}).  Thus, with the choice of the
coefficients $a,b$ given by Eqs.~(\ref{a},\ref{b}), the theory defined
by the action density (\ref{Lscalar}) is equivalent to the scalar QED.

Before we close this section, we make a brief comment on the relation of
the above model to the ${\rm CP}^{N-1}$ model.\cite{AA} (See also
Chapter 5 of ref.~\cite{BKY}.)  The ${\rm CP}^{N-1}$ model can be
obtained formally from the action density (\ref{Lscalar}) by imposing a
non-linear constraint
\be
\phi^{I*} \phi^I = {N v^2 \over 2}.
\ee
Since the $O(2N)$ non-linear sigma model is equivalent to the $O(2N)$
linear sigma model, the ${\rm CP}^{N-1}$ model is equivalent to the
model discussed in this section, and therefore to the scalar QED.

\section{Conclusion}

We have seen that once the mass parameters are fine-tuned, theories
which are non-renormalizable by the usual power counting rule reduce to
renormalizable theories at energies below the cutoff $\Lambda$ as long
as we ignore quantities inversely proportional to $\Lambda$. The
renormalized parameters at a finite energy scale $\mu$ appear as the
marginally irrelevant dependence on the cutoff of the order of ${1 \over
\ln {\Lambda \over \mu}}$.  We have studied two matter-only models
without elementary gauge fields whose current-current contact
interactions at the cutoff scale give rise to interactions mediated by
the abelian gauge fields at low energies.  The gauge field was
dynamically generated to complete renormalizability of the theory just
as the sigma field is dynamically generated in the $O(N)$ non-linear
sigma model.

In this paper the massive photon was obtained not as a consequence of
the Higgs mechanism: it appeared as part of the gauge fixing terms.  The
mass is simply allowed by the Ward identities in the case of abelian
gauge theories.  In the next paper we will study a matter-only model
which exhibits the Higgs mechanism.\cite{next}

It should be interesting to extend our analysis to non-abelian gauge
theories.  We expect that the covariant gauge fixing term will arise
naturally just as for QED, but in the case of non-abelian gauge theories
the covariant gauge requires the Faddeev-Popov (FP) ghosts.  Hence, for
a matter-only model to become a non-abelian gauge theory, the FP ghosts
must be generated dynamically.  It will be extremely interesting if this
is the case.

In ref.~\cite{KM} an induced QCD was studied in the $1/N$ expansions in
the hope of uncovering the non-perturbative dynamics of QCD.  We also
hope that the reformulation of gauge theories as matter-only theories
will find practical applications in understanding, for example, the
physics of QED at energies very high compared to the electron mass but
still much lower than the cutoff scale.

\vspace{0.5cm} This work was partially supported by the Grant-In-Aid for
Scientific Research (No.~11640279) from the Ministry of Education,
Science, and Culture, Japan.

\appendix

\section{Ward identities}

In determining the coefficients of the counterterms, we have imposed Ward 
identities.  We remind the reader of the Ward identities for both QED
with electrons and the scalar QED.  For QED with electrons, we have
three Ward identities to satisfy:
\ba
k_\mu \Pi_{\mu\nu} (k) &=& {k_\nu \over e^2} \left( m_\gamma^2 + {1
\over \xi} k^2 \right)\\ - i k_\mu \Pi_\mu^{\psi\psibar} (p,k) &=& {i
\over \sqrt{N}} \left( \Pi^{\psi\psibar} (p) - \Pi^{\psi\psibar} (p+k)
\right)\\ k_\alpha \Pi_{\alpha\beta\gamma\delta} &=& 0,
\ea
where $\Pi_\mu^{\psi\psibar} (p,k)$ is the electron-photon interaction
vertex, and $\Pi^{\psi\psibar} (p)$ is the inverse electron propagator
with momentum $p$.  
\begin{center}
\epsfig{file=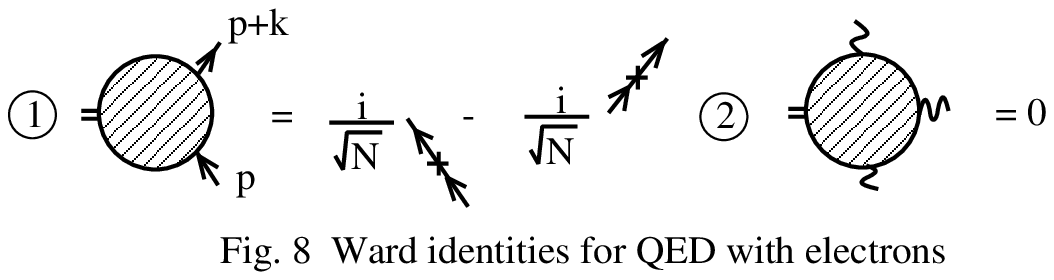,height=3cm}
\end{center}

The Ward identities for the scalar QED are similarly given by
\ba
k_\mu \Pi_{\mu\nu} (k) &=& {k_\nu \over e^2} \left( m_\gamma^2 + {1
\over \xi} k^2 \right)\\ - i k_\mu \Pi_\mu^{\phi\phi^*} (p,k) &=& {i
\over \sqrt{N}} \left ( \Pi^{\phi\phi^*} (p) - \Pi^{\phi\phi^*}
(p+k)\right)\\ - i k_\mu \Pi_{\mu\nu}^{\phi\phi^*} (p,k,l) &=& {i \over
\sqrt{N}}\left( - \Pi_{\nu}^{\phi\phi^*} (p,l) + \Pi_{\nu}^{\phi\phi^*}
(p+k,l) \right)\\ k_\alpha \Pi_{\alpha\beta\gamma\delta} &=& 0
\ea
using a similar notation as for QED.
\begin{center}
\epsfig{file=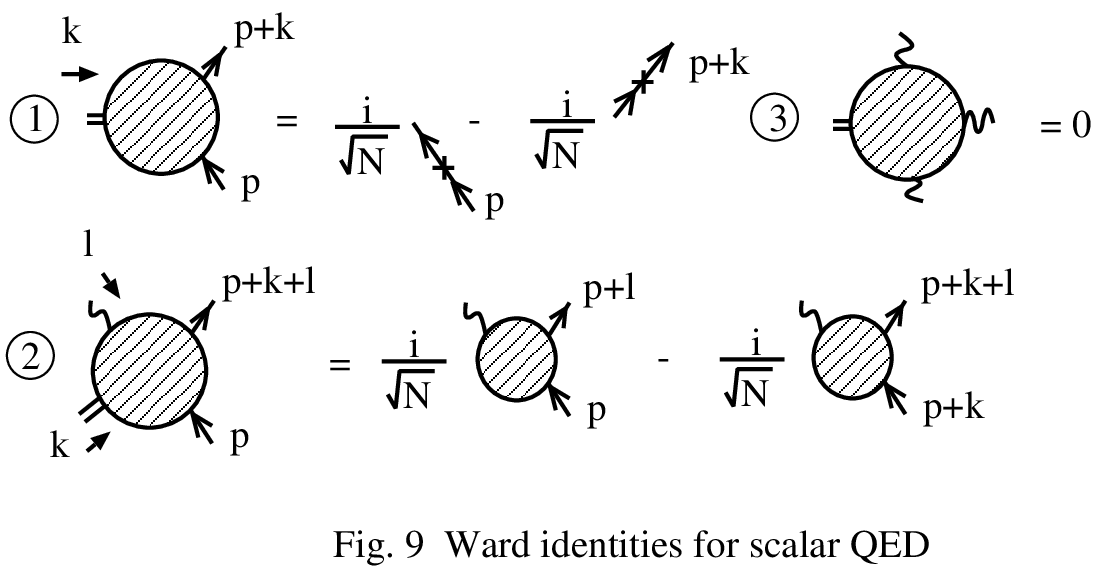,height=5cm}
\end{center}

\section{Integrals with a momentum cutoff}

In calculating the one-loop integrals with a momentum cutoff, we have
used the following formulas where contributions of order ${m^2 \over
\Lambda^2}$ or less are ignored.
\ba
\int_{p<\Lambda} {1 \over p^2 + m^2} &\equiv& \int_{p^2 < \Lambda^2}
{d^4 p \over (2\pi)^4} {1 \over p^2 + m^2} \nonumber\\
&=& {1 \over (4\pi)^2} \left(
\Lambda^2 - m^2 \ln {\Lambda^2 \over m^2}\right)\\ \int_{p < \Lambda}
{p_\mu p_\nu \over (p^2 + m^2)^2} &=& {\delta_{\mu\nu} \over 4} \int_p
{p^2 \over (p^2+m^2)^2} \nonumber\\ &=& {\delta_{\mu\nu} \over 4} {1
\over (4\pi)^2} \left( \Lambda^2 - 2 m^2 \ln {\Lambda^2 \over m^2} + m^2
\right)\\
\int_{p<\Lambda} {1 \over (p^2 + m^2)^2} &=& {1 \over (4\pi)^2} \left
( \ln {\Lambda^2 \over m^2} - 1 \right)\\
\int_{p<\Lambda} {p_\mu p_\nu \over (p^2 + m^2)^3} &=& {\delta_{\mu\nu}
\over 4} {1 \over (4 \pi)^2} \left( \ln {\Lambda^2 \over m^2} - {3 \over 
2} \right)
\ea

\end{document}